\documentclass[prb]{revtex4}

\usepackage{graphicx}

\begin{document}

\title{ Bohr's atomic model revisited}
\author{F. Caruso}
\affiliation{Centro Brasileiro de Pesquisas F\'{\i}sicas \\
Rua Dr. Xavier Sigaud 150, 22290-180, Rio de Janeiro, RJ, Brazil}

\email{francisco.caruso@gmail.com}

\author{V. Oguri}
\affiliation{Instituto de F\'{\i}sica Armando Dias Tavares da Universidade do Estado do
Rio de Janeiro \\
Rua S\~ao Francisco Xavier 524, 20550-013, Rio de Janeiro, RJ, Brazil}

\email{oguri@uerj.br}

\begin{abstract}{Bohr's atomic model, its relationship to the radiation spectrum of the hydrogen atom and the inherent hypotheses are revisited. It is argued that Bohr could have adopted a different approach, focusing his analyzes on the stationary orbit of the electron and its decomposition on two harmonic oscillators and then imposing, as actually he did, Planck's quantization for the oscillators' energies. Some consequences of this procedure are examined.}
\end{abstract}

\maketitle

\section{Introduction}

Niels Bohr, in 1913, set forth that the elementary quantum of action -- Planck's constant --, originally introduced for explaining the emission of light from a black-body, should also be necessary to ensure matter stability.\cite{Niels} His atomic model was probably one of the most intriguing and fruitful conception of the inner structure of chemical elements. Moreover, Dirac has manifested his belief ``{\it that the introduction of these ideas by Bohr was the greatest step of all in the development of quantum mechanics.}".\cite{Dirac} The solution of the dynamical instability of Rutherford's atom proposed by Bohr was in direct contradiction with what was expected from Maxwell-Lorentz theory of accelerated charges, enlarging the crisis of Classical Physics (extending it to matter) which had started some years before by the two seminal papers by Planck and Einstein which shed light on both Lord Kelvin's clouds. In Bohr's own words, ``{\it {\rm [it]} seems to be a general acknowledgment of the inadequacy of the classical electrodynamics in describing the behavior of systems of atomic size.}"  However, Bohr's atomic model was very successful in describing the regularity of atomic spectra, specially the regularities of Balmer's spectrum for hydrogen atoms. Referring to this success, although not yet comprehensible, James Jeans very quickly admitted in public how ingenious were Bohr's ideas\cite{Rosen}  and Einstein said that there should be something behind his result since it is very hard to believe that Rydberg constant could be correctly obtained by chance.\cite{Franc} Indeed, spectroscopy gave the empirical basis for the development of Bohr's atomic model and, why not to say, to the development of quantum theory. As Scott Tanona has stressed,\cite{Scott}  the Danish physicist
\begin{quotation}
\noindent ``{\it attempted to use well-defined phenomena as a secure starting point for the development of theory and searched for the most general features that any atomic theory was to have in the face of the novel introduction of quantum effects.}"
\end{quotation}


It was in his paper ``On the constitution of atoms and molecules''\cite{Niels} that Bohr, still using Newtonian Mechanics, set the basis for a new quantum description of matter, by disclosing the inadequacy of classical electromagnetism to describe Atomic Physics.  His work had the merit of stimulating many physicists to think about the necessity of elaborating a new theoretical description of the microphysics. It is curious to note that it was Newtonian Mechanics -- not Electromagnetism --  which was first replaced by a Quantum Theory (Quantum Mechanics). Indeed, Bohr had a great influence on Heisenberg and Louis de Broglie works and it is well known how Schr\"odinger's early works were influenced by de Broglie's wave-particle duality ideas. However, Quantum Electrodynamics still had to wait some decades to be established. It was also in this paper\cite{Niels} where the {\it correspondence principle} was introduced for the first time. It is commonly accepted that this principle is a very powerful heuristic argument that has played an important r\^ole on Bohr's starting point of view on Quantum Physics,  by requiring that the classical results should follow from the quantum results in the limit of large quantum numbers. Although counterexamples have shown that this idea is not generally true,\cite{Liboff} from the epistemological point of view, we rather agree with Tanona when he says\cite{Scott} that
\begin{quotation}
\noindent ``{\it For Bohr, the correspondence principle was not merely a heuristic tool for the development of quantum theory; nor was it {\rm primarily} concerned with the asymptotic agreement between classical and quantum theory. Rather, its main purpose within Bohr's empirical approach was to bridge the epistemological gap between empirical phenomena and the unknown atomic structure, which is responsible for the phenomena, by associating classical properties of spectra with atomic properties.}
\end{quotation}

Within a difference of six to seven months, Bohr introduced two new postulates, in order to save mechanical stability of atoms as reproduced here:

\begin{description}
\item[ ] (i) An atomic system based on Rutherford's model can only exist in certain {\it stationary states} (orbits) having energies
\begin{equation}
\{\epsilon_1, \epsilon_2, \epsilon_3, \dots \}
\end{equation}
and can be partially described by the laws of Classical Mechanics. Subsequently, Bohr called this ``{\it the quantum postulate}";

\item[ ] (ii) The {\it emission} (or {\it absorption}) of electromagnetic radiation occurs only due to transitions between two stationary orbits, in such a way that the frequency ($\nu$) of the emitted (or absorbed) radiation is given by

\begin{equation}\label{delta_E}
\nu = \frac{\big| \epsilon_f - \epsilon_i \big|}{h}
\end{equation}
where $h$ is Planck's constant and $\epsilon_f$ and $\epsilon_i$ are,
respectively, the energy values of the two orbits involved in the transition. In other words, the energy
($\epsilon$) of the emitted or absorbed photon is given by Planck's law
 $$ \epsilon = h\nu$$
 \end{description}

This is almost always what is frequently taught in a Modern Physics course. However, there are, indeed, many other assumptions hidden in Bohr's model. A complete list was clearly settled by Sir Edmund Whittaker:\cite{Whitta}

\begin{description}
\item[] 1) ``atoms produce spectral lines one at a time;
\item[] 2)  that a single electron is the agent of the process (...);
\item[] 3)  the Rutherford atom provides a satisfactory basis for exact calculations of wave-lengths of spectral lines;
\item[] 4) the production of atomic spectra is a quantum phenomenon;
\item[] 5) an atom of a given chemical element may exist in different {\it states}, characterized by certain discrete values of its angular momentum and also discrete values of energy (...);
\item[] 6) that in quantum-theory, angular momenta must be whole multiples of $\hbar$ (...);\cite{This}
\item[] 7) that {\it two} distinct states of the atom are concerned in the production of a spectral line (...);

\item[] 8) that the Planck-Einstein equation $E = h \nu$ connecting energy and radiation-frequency holds for {\it emission} as well as absorption."

\end{description}

To this list, Whittaker adds a last postulate which, due to its epistemological relevance, should be detached, namely the principle that {\it we must renounce all attempts to visualize or to explain classically the behavior of the active electron during a transition of the atom from one stationary state to another}. This renouncement is exactly what Bohr assumed when he attempted to build up a quantum model (mechanically stable) for the atom with an eye kept on his electromagnetic discrete spectrum and another kept on the single electron agent of the process, without having a real understanding of the quantum process actually involved in radiation emission. In this sense, the {\it correspondence principle} is much more the epistemological link between the observed electromagnetic spectra and the inferred mechanical orbital motion of the electron than a general connection between classical and quantum Physics.

For his project to succeed, there is still another assumption (not mentioned by Whittaker) that Bohr should have made in the first paper\cite{Niels} of his famous trilogy which often deserves no attention, although it seems {\it quite strange}. We are talking about the way Bohr saw the mechanism of binding an electron to a nucleus and how he has related its revolution frequency ($f$) to the frequency ($\nu$) of the emitted homogeneous radiation, as will be seen in the next Section. This was an important open question at that time.

\section{Binding electrons to positive nucleus}\label{balmer-bohr}

The idea that one should abandon classical concepts to understand atomic phenomena was almost clear to Bohr by the time of his doctorate thesis. By the end of 1911, he wrote a letter to McLaren expressing his conviction that very little was known from the movement of electrons in metals. In his dissertation it was already evident for him that Lorentz electron theory was not enough to surmount some unexplained phenomena. He guessed there was something wrong about how electrons interact with ions. In a couple of years, Bohr moved his attention from {\it free} electrons (as in metals) to {\it bound} electrons.

Following L\'eon Rosenfeld's testimony,\cite{Rosen} by July 1912, in a memorandum addressed to Rutherford, Bohr already expresses the kinetic energy of an electron inside the atom as proportional to its revolution frequency, but without fixing the proportionality constant neither relating it to Planck's constant. In this memorandum nothing else is said about this energy. Rosenfeld seems to accept that in 1912 Bohr already considered the argument he used in the 1913 first paper, namely: that the bound energy of the electron is equal in magnitude to half the mean energy in the revolution period. Let us briefly review the steps.

Starting from Rutherford's  hydrogen model, in which the atom is made of
a nucleus with very tiny dimensions of positive electric charge ($+e$), and of an electron with negative charge ($-e$), Bohr admitted the electron to describe elliptical stationary orbits with velocity ($v$) much less than that of light in the vacuum ($c$),
and that there is no radiation loss. In addition, the interaction between the electron and the nucleus could be described by an electrostatic Coulombian force ($F$) given by
$$ F = - \frac{Ze^2}{r^2}$$
where $r$ is the relative distance.

In a first moment of the 1913 famous paper,\cite{Niels} Bohr focused his attention on how an electron can be bound to a nucleus. In the second section of his paper, we read:
\begin{quotation}
\noindent
``{\it {\rm (...)} let us assume that the electron at the beginning of the interaction with the nucleus was at a great distance apart from the nucleus, and had no sensible velocity relative to the latter. Let us further assume that the electron after the interaction has taken place has settled down in a stationary orbit around the nucleus.}"
\end {quotation}

\noindent And he goes on saying that ``{\it let us now assume that, during the binding of the electron, a homogeneous radiation is emitted of a frequency $\nu$, equal to half the frequency of revolution of the electron in its final orbit}".

In our notation, this means to admit the {\it frequency postulate}, namely that
\begin{equation}\label{freq_postulate}
\nu = \frac{1}{2}\, f
\end{equation}

The relations between the electron revolution frequency $(f)$ of mass $m$ and the ellipse major semi-axis ($a$) for a given energy ($\epsilon$) are given by\cite{Niels}
 \begin{equation} \label{rel_class}
 \left\{
\begin{array}{l} \displaystyle
f = \frac{1}{\pi Z e^2} \sqrt{\frac{2|\epsilon|^3}{m}}  \\
\ \\
\displaystyle
2a = \frac{Z e^2}{|\epsilon|}
\end{array}
\right.
\end{equation}

When there is no restriction to the values of energy, frequency or the ellipse's axes, the only restriction to those values are those imposed from the previous relations. In addition, classically, when the motion of an electron in a particular orbit (with definite frequency) is considered to be the cause of energy emission from the atom, it should necessarily follows that light with different frequencies -- corresponding to each one of the Fourier components of its motion -- would be emitted. This is, of course, not what was observed.

Thus, according to postulate (i), the energy value set for stationary orbits is a discrete set, namely,
$$\{\epsilon_n\}  \qquad \qquad (n=1,2,\dots)  $$

The relationship between electron orbital movement and the emitted frequency should be determined, according to Bohr, by the correspondence principle. In fact, in another point of the paper, Bohr admitted that the energy of each state depends on the revolution frequency, which imposes a new condition between the energy and the revolution frequency of a stationary state

\begin{equation} \label{Bohr_hip1}
|\epsilon_n| = h f_n g(n)
 \end{equation}
where $g(n)$ is an unknown function. His idea was using his correspondence principle to fix the functional dependence of $g(n)$ in order to correctly reproduce Balmer's formula for the hydrogen spectrum.

Replacing the frequency given by equation~(\ref{Bohr_hip1})
in the corresponding equation~(\ref{rel_class}), it follows that
\begin{equation}
\displaystyle
 \frac{|\epsilon_n|}{h g(n)} =
 \frac{1}{\pi Z e^2} \sqrt{\frac{2|\epsilon_n|^3}{m}}
\end{equation}
which implies that
\begin{equation}\left\{
\begin{array}{l} \displaystyle
|\epsilon_n| =  \frac{\pi^2 m Z^2 e^4}{2 h^2} \ \frac{1}{g^2(n)}  \\
\ \\
\displaystyle
f_n =  \frac{\pi^2 m Z^2 e^4}{2 h^3} \ \frac{1}{g^3(n)}
\end{array}
\right.
\end{equation}

According to postulate (ii), having emission or absorption due to the transition between states of energies
$\epsilon_n$ e $\epsilon_l$, the frequency ($\nu$) of the emitted or absorbed radiation is given by
\begin{equation}\label{Bohr_ryd}
\nu =  \nu_{ln} = \frac{\big| \epsilon_l - \epsilon_n \big|}{h} =
 \frac{\pi^2 m Z^2 e^4}{2 h^3} \left| \frac{1}{g^2(l)} - \frac{1}{g^2(n)}
 \right|
 \end{equation}
Note that the frequency of the radiation is in principle different from the revolution frequency.

For the hydrogen atom ($Z=1$), equation~(\ref{Bohr_ryd}) is compatible with the Rydberg formula, written as
\begin{equation}
\displaystyle
\nu = c R_H \ \left(\frac{1}{l^2} - \frac{1}{n^2}\right)
\end{equation}
if the function  $ g(n) = b n $, where $b$ is a constant to be fixed.

Therefore, we can write
\begin{equation} \left\{
\begin{array}{l} \displaystyle
|\epsilon_n| =  \frac{\pi^2 m e^4}{2 h^2} \  \frac{1}{b^2 n^2}  \\
\ \\
\displaystyle
f_n =  \frac{\pi^2 m e^4}{2 h^3} \  \frac{1}{b^3 n^3}
\end{array}
\right.
\end{equation}

The $b$ constant can be determined considering the transition between two states with neighboring energy values
$\epsilon_n$ and $\epsilon_l$, so that $n=l+1$, in the limit of large values of  $n$. In this limit, $f_n = f_l$, and Bohr considers  that the frequency ($\nu_{ln}$), Eq.~(\ref{Bohr_ryd}), of the emitted radiation should be equal to the electron revolution frequency ($f_n$).
This hypothesis is known as Bohr's {\it frequency correspondence principle}.

Thus,
\begin{equation}
 \displaystyle  \lim_{n \to \infty} \nu_{ln} =
 \frac{\pi^2 m e^4}{2 h^3} \ \frac{1}{b^2 n^2}
 \left|  \left(\frac{n}{l} \right)^2 - 1 \right|
\end{equation}

For very large value of $l$, $l=n$, the limit becomes
\begin{equation}
 \displaystyle  \lim_{n \to \infty} \nu_{ln}
=\frac{\pi^2 m e^4}{2 h^3} \ \frac{2}{b^2 n^3}
\end{equation}
and, according to the correspondence principle,
\begin{equation}
\lim_{n \to \infty} \nu_{ln} = \lim_{n \to \infty} f_n \quad \Longrightarrow \quad
b=1/2
\end{equation}

In this case, the initial Bohr's hypothesis, equation~(\ref{Bohr_hip1}), can be expressed by
\begin{equation}\label{Bohr_eq_freq}
\fbox{$\displaystyle  |\epsilon_n| = \frac{n h f_n}{2}$}
\end{equation}

To this result Bohr has dedicated a comment concerning the {\it ad hoc} relation between the frequency of the emitted radiation and that of electron revolution he has assumed at the beginning of the paper and its interpretation. Essentially he says that ``{\it from this assumption we get exactly the same expressions as before for stationary states, and from these by help of the principal assumptions on p.~7 the same expression for the law of the hydrogen spectrum {\rm [see below]}}". Without being able to establish a more clear interpretation of this result he concludes the remark by saying that the preliminary considerations, implicit in the above derivation, should be seen ``{\it only as a simple form of representing the results of the theory}".

The atomic energy spectrum is given by
\begin{equation}\label{Bohr_eq_espec}
 \epsilon_n = -  \left( \frac{2 \pi^2 m e^4}{h^2} \right) \ \frac{1}{n^2}
 \qquad \qquad (n=1,2,\dots)
\end{equation}

\noindent
From postulate (ii),  he had obtained for the emitted or absorbed radiation Ritz's formula:

\begin{equation}\label{freq_transicao}
 \nu_{ln} = c R_{_H}  \left| \frac{1}{l^2} - \frac{1}{n^2} \right|
\end{equation}
where $R_{_H}$ is the Rydberg constant. Fixing $l=2$ and letting $n$ vary, one gets Balmer's formula.

Therefore, the frequency of the emitted radiation during the transition between the orbit labeled by $n$ and the next one is equal to the electron revolution frequency in the $n$-orbit, just for large values of $n$. It is important to remember that many of Bohr's predecessors as Lorentz, Zeeman, Larmor and Thomson, contrary to him, used in their classical calculations of the radiation frequency from the movement of atomic constituents, the misleading assumption that this frequency was equal to that of electron revolution. This was proved to be wrong for the lower energy levels.
What Bohr has primarily shown, based on the hydrogen spectra data, is that there is an asymptotic limit of the quantum result which coincides with the classical result just in the limit of large $n$ values.

\section{Energy levels and angular momentum quantization}

After having deduced Balmer's formula from the postulate of quantization of the emitted energy and from the correspondence principle,
Bohr alternatively assumed, like Haas, that an electron of charge $-e$ and mass $m$ describes a circular orbit of radius $r$  under the the Coulombian attractive force of the positive nucleus (consider hydrogen for simplicity), such that
\begin{equation}\label{forcas_iguais}
\frac{mv^2}{r} = \frac{e^2}{r^2}  \qquad \Longrightarrow \qquad
m v^2 = \frac{e^2}{r} \qquad \Longrightarrow \qquad
(mvr)^2 = m e^2 r
\end{equation}
where $v$ is the electron velocity.

The orbital angular momentum ($L$)  of the electron in respect to the nucleus is given by
\begin{equation}
 L = mvr = m\omega r^2
\end{equation}
The radius of the orbit and the energy could be written as a function of angular momentum as
\begin{equation}
\left\{ \begin{array} {l}
\displaystyle
r = \frac{L^2}{me^2} \\
\ \\
\displaystyle
 \epsilon = \frac{1}{2} m v^2 - \frac{e^2}{r}  = - \frac{e^2}{2r} =
- \frac{m e^4}{2 L^2}
\end{array}
\right.
\end{equation}

From the classical point of view, since angular momentum can vary continuously, any orbit is a possible orbit.
There will be no reason in this case that the fundamental state should have a particular radius.
However, assuming a new postulate -- the quantization of angular momentum --
\begin{equation}
 L = n \hbar \qquad \quad (n=1,2,3,...)
\end{equation}
we get a class of possible radius $r_n$ for each orbit $n$,
\begin{equation}
r_n = n^2 \frac{\hbar^2}{me^2}
\end{equation}
where $\hbar = h/(2\pi)$.

Analogously, the total energy can be written as
\begin{equation}\label{energ_bohr_Z}
\epsilon_n = - \frac{1}{2} \ \frac{m e^2}{\hbar^2}
\frac{e^2}{n^2} = - \frac{e^2}{2a} \frac{1}{n^2}
\end{equation}

In this way, admitting circular orbits for the electron in the stationary states, the correspondence principle can be substituted by the quantization of angular momentum postulate, given rise to the energy quantization and to Balmer's formula.
So, angular momentum quantization leads to energy quantization in an atomic transition.

\section{Bohr's atomic electron as an harmonic oscillator}\label{Bohr_sec_osc}

In 1920, Bohr wrote:\cite{Bohr2}
\begin{quotation}
\noindent ``{\it On account of the general correspondence between the spectrum of an atom and the decomposition of its motions into harmonic components, we are led to compare the radiation emitted during the transition between two stationary states with the radiation which would be emitted by a harmonically oscillating electron on the basis of the classical electrodynamics.}"
\end{quotation}

According to Bohr's circular model, the energy of an electron of mass $m$
moving with frequency $\omega=2\pi f$, in an orbit with radius $r_{_0}=a$  and
velocity $v_{_0}=\omega r_{_0}$, can be written as
\begin{equation}\label{mod_energia_bohr}
\left| \epsilon_{\mbox{\tiny Bohr}} \right| =  \frac{\omega L}{2} = \frac{1}{2}m \omega^2 r^2 = \frac{1}{2} m v^2_{_0}
\end{equation}
where $L = m v_{_0} r_{_0}$ is the orbital angular momentum of the electron in respect to the nucleus.

Moreover, a particle of mass $m$, subject to a central force of the type $\vec F = - m \omega^2  \vec r$,
oscillate with a frequency $\omega = 2\pi f$.
The general solution for the movement of this oscillator, at any instant $t$,
is given by
\begin{equation}
\vec r = \vec r_{_0} \cos \omega t \ + \
\frac{\vec v_{_0}}  {\omega} \, \mbox{sen} \, \omega t
\end{equation}
where $\vec r_{_0}$ and $\vec v_{_0}$ are, respectively, the initial position and velocity of the particle, and
$\vec r $ the position at time $t$.

Supposing that $\omega$, $\vec r_{_0}$ and $\vec v_{_0}$ have the same values that their correspondents in Bohr's circular model, the total energy of the oscillator is given by

\begin{equation}
\epsilon_{\mbox{\tiny osc}} = m \omega^2 r^2_{_0}
\end{equation}

Comparing this result with the energy of Bohr's model, we obtain

\begin{equation}
\left| \epsilon_{\mbox{\tiny Bohr}} \right| =
\frac{\epsilon_{\mbox{\tiny osc}}}{2}  \qquad \Longrightarrow \qquad
\omega L =   \epsilon_{\mbox{\tiny osc}}
\end{equation}

Therefore, if the energy of the oscillator satisfies Planck's quantization rule,
$ \epsilon_{_{\mbox{\tiny osc}}} = n h f$,
the  energy of  Bohr's circular model, postulated by Eq.~(\ref {Bohr_hip1}), can be written --
comparing to Eq.~(\ref{mod_energia_bohr})
-- as
\begin{equation}
\epsilon_{\mbox{\tiny Bohr}} = \frac{n h f}{2}
\end{equation}
and, consequently, the angular momentum is quantized:
\begin{equation}
L=n\hbar \qquad \quad (n=1,2,3,...)
\end{equation}

\section{An alternative approach}\label{Bohr_altern}

Taking into account all we have just reviewed, one can now argue whether this scenario was at that time indeed unique or, at least, whether it was really the ``simplest" one. Could Bohr avoid anyone of his postulates? Could Bohr had achieved the same results without introducing the {\it frequency correspondence principle} -- a postulate which is ultimately connected to the unusual assumption that the frequency of the emitted radiation is equal to half the frequency of revolution of the atomic electron in its final orbit? The answer seems to be {\it yes}, but then, why Bohr did not adopt it?

It will be shown in this section that he could have thought in a different way, and a sketch of why he might not will be presented.  In other words, so far as the problem of the electron orbit was concerned, it was possible at that time to exploit further the classical results including only two main postulates: electrons have stationary orbits inside the atom ({\it the quantum postulate}); and the energy of {\it any} harmonic oscillator is given by Planck's quantization law.

\begin{figure}[htbp]
\centerline{\includegraphics[width=5cm]{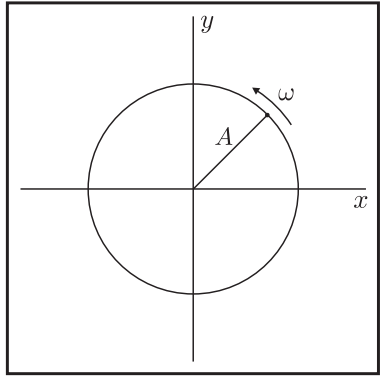}}
\caption{Circular movement of a particle with frequency $\omega$ and radius $A$.}
\label{bohr}
\end{figure}

Consider the classical circular movement of a charge in a plane with frequency $\omega = 2\pi f$, as shown in figure~\ref{bohr}.

The circular movement is characterized by an acceleration $a = \omega^2 r$. It was well known from acoustic that this movement can be thought as the superposition of two harmonic simple oscillators mutually perpendiculars with amplitudes

 \begin{equation} \label{harm_comp}
 \left\{
\begin{array}{l} \displaystyle
x = A \cos (\omega t + \theta)  \\
\ \\
\displaystyle
y = A \sin (\omega t + \theta)  \\
\end{array}
\right.
\end{equation}

\noindent and acceleration components

\begin{equation} \label{acel_comp}
 \left\{
\begin{array}{l} \displaystyle
a_x = -\omega^2 A = - \omega^2 A \cos (\omega t + \theta)  \\
\ \\
\displaystyle
a_y = -\omega^2 A = -\omega^2 A \sin(\omega t + \theta)  \\
\end{array}
\right.
\end{equation}

The potential energy of one oscillator (in $x$-direction) is
\begin{equation}
 E_p = \frac{kx^2}{2} = \frac{kA^2}{2} \cos^2 (\omega t + \theta)
 \end{equation}
with $k = m\omega^2$, and the kinetics energy is
\begin{equation}
E_k = \frac{1}{2} mv^2 = \frac{1}{2} k A^2 \sin^2 (\omega t + \theta)
\end{equation}

Thus, the total energy of each oscillator ($i=x, y$) is
\begin{equation}
E_i = E_p + E_k = \frac{1}{2} k A^2
\end{equation}

\noindent and the total energy is
\begin{equation}
E = E_x + E_y = k A^2
\end{equation}

By other side, it is well known that when the circular movement of a massive and charged particle is due to a central Coulombian attractive force, its radius $A$ can be expressed as a function of the angular momentum and other constants associated to the particle (mass and charge) like
\begin{equation}
A = \frac{L^2}{me^2}
\end{equation}

Squaring this radius and then substituting its value and $k = m\omega^2$ in the total energy we get
\begin{equation}\label{form_energ}
E = \frac{\omega^2 L^4}{me^4}
\end{equation}

The angular frequency can also be expressed in terms of the angular momentum by
\begin{equation}\label{form_freq}
\omega = \frac{m e^4}{L^3}
\end{equation}

Therefore, from Eqs.~(\ref{form_energ}) and (\ref{form_freq}), a relationship between energy and angular momentum is obtained:
\begin{equation}\label{energy_l}
E = \frac{m e^4}{L^2} = L \omega
\end{equation}

This is strictly a classical result. At this point we can introduce Planck's quantization for the energy of both simple harmonic oscillators, $E = n\hbar \omega$, obtaining

\begin{equation}
E = n_1\hbar \omega + n_2\hbar \omega = (n_1 + n_2) \hbar \omega \qquad \Rightarrow \qquad \fbox{$\displaystyle E = n \hbar \omega$}
\end{equation}

\noindent and, from Eq.~(\ref{energy_l}), the quantization of orbital angular momentum

\begin{equation}
\fbox{$\displaystyle L= n \hbar$}
\end{equation}

Note that this result do not depend on {\it any} kind of consideration about the electron transition between two orbits. We are just treating the circular motion of {\it one stationary orbit} as a formal composition of two simple harmonic oscillators of the same frequency and postulating that the energy of {\it any} harmonic oscillator is quantized according to Planck's quantization law. It seems to us that this way of treating the problem of electron movement in Bohr's model is at least more ``economic": the frequency postulate, given by Eq.~(\ref{freq_postulate}), could therefore be avoided.

Now the question: Why Bohr did not consider this possibility? First of all just because he did not take the concept of {\it classical orbits} very seriously. Indeed, as Folse remembers,\cite{Folse}
\begin{quotation}
\noindent ``{\it The now well-known ``electron orbits" were simply a pseudo-classical means of representing these stationary states. Though the degree of seriousness with which Bohr took this model waxed and waned throughout this period, certainly he was only rarely tempted to take the orbiting electron as a literal spatio-temporal description of the physical situation inside the atomic system. Furthermore, he was alarmed by the tendency of many other physicists to do so.}"
\end{quotation}

In a certain sense, we can argue that there is a second stronger historical reason. The fact that Bohr was not able to realize how to introduce Planck's constant in the description of matter until he developed his theory of atomic structure to include not {\it one} but a {\it series} of stationary states, between which atomic transitions were {\it discontinuous} (Section~2). Such discontinuity was essential to understand the Balmer formula, at least qualitatively. The clue relating theory to experiment, as discussed in this paper, is the {\it frequency correspondence principle}, which has led to a correct quantitative description of the observed spectra of hydrogen atoms. Therefore, Bohr's perception that the atomic system should change its state discontinuously is a necessary consequence of his belief that the only allowed transitions are between {\it stationary states}. {\it Two} -- not just {\it one} -- states are needed (point 7 of Whittaker's list). At this point is quite unavoidable not to remember the Dirac's comment\cite{PAM} concerning the main point of Heisenberg's 1925 idea that theory should concentrate on observed quantities:
\begin{quotation}
\noindent ``{\it  Now, the things you observe are only very remotely connected with the Bohr orbits. So Heisenberg said that the Bohr orbits are not very important. The things that are observed, or which are connected closely with the observed quantities, are all associated with two Bohr orbits and not with just one Bohr orbit: {\it two} instead of {\it one}.}"
\end{quotation}

The aforementioned discussion suggests that such Heisenberg's point of view was not so strange to Bohr twelve years later, when he based his atomic model on two hypotheses: the {\it frequency postulate} and the {\it frequency correspondence principle}. His choice seems to be much more dictated from philosophical than physical reasons. The reasoning presented in this Section is an alternative way to establish the quantization of both the energy and the angular momentum of the electron in a {\it stationary state} without going through more states. Perhaps, an exploitation of this reasoning at that time could drive to a version of the Wilson-Sommerfeld quantization rule, as suggested in the first talk Bohr gave in honor of C.~Christiansen, published in 1918.\cite{Bohr2} Indeed, in a nutshell, it was in this paper that Bohr had shown the now well known result that Planck's quantization law for the energy of an one-dimensional harmonic oscillator is equivalent to the condition
\begin{equation}\label{somm}
\oint p \mbox{d} q = n h
\end{equation}
where the integral is taken over a complete oscillation of the variable $q$ between its limits and $p$ is the canonically conjugated momentum.

Bohr arrived to this conclusion after he had realized how the {\it adiabatic hypothesis}, introduced by Ehrenfest\cite{Paul} and called by the former the {\it principle of mechanical transformability}, could give support to his definition of {\it stationary orbits} or, in other words, how he could justify to fix a number of the atomic states {\it among the continuous multitude of mechanically possible motions}. This principle, as Bohr mentioned in 1918,

\begin{quotation}
\noindent ``{\it allows us to overcome a fundamental difficulty  which at first sight would seem to be involved in the definition of the energy difference between two stationary states which enters in the relation {\rm [}$E^\prime - E^{\prime\prime} = h\nu${\rm ]}. In fact we have assumed that the direct transition between two such states cannot be described by ordinary  mechanics, while on the other hand we possess no means of defining an energy difference between two states if there exists no possibility for a continuous mechanical connection between them.}"
\end{quotation}

We have shown in Section~\ref{Bohr_altern} that there is an alternative through which it is possible to fix not only the energy for the transition but rather to fix it and the angular momentum for each stationary orbit.

\begin{acknowledgments}
We would like to thank S\'ergio Joffily, Alfredo Marques and H\'elio da Motta for useful remarks.
\end{acknowledgments}


\begin{references}
\bibitem{This} This should actually not be considered an hypothesis, as is pointed out in this paper.
\end{references}
\end{document}